\documentclass[letter,
               keeplastbox,   
               ]{jacow}
\usepackage{pdfpages,multirow,ragged2e} %
\makeatletter%
	\ifboolexpr{bool{xetex}}
	 {\renewcommand{\Gin@extensions}{.pdf,%
	                    .png,.jpg,.bmp,.pict,.tif,.psd,.mac,.sga,.tga,.gif,%
	                    .eps,.ps,%
	                    }}{}
\makeatother

\ifboolexpr{bool{xetex} or bool{luatex}} 
 {}                                      
 {\usepackage[utf8]{inputenc}}           

\usepackage[USenglish]{babel}

\ifboolexpr{bool{jacowbiblatex}}%
 {%
  \addbibresource{jacow-test.bib}
  \addbibresource{biblatex-examples.bib}
 }{}
\begin{document}
\def\Up{\boldsymbol{\uparrow\hspace{-5.5pt}\uparrow}}
\def\Dn{\boldsymbol{\downarrow\hspace{-5.5pt}\downarrow}}

\title{Foundations of Iterative Learning Control }

\author{S. R. Koscielniak\thanks{shane@triumf.ca}, TRIUMF, Vancouver, Canada}
	
\maketitle

\begin{abstract}
Iterative Learning Control (ILC) is a technique for adaptive feed-forward control of electro-mechanical plant that either performs programmed periodic behavior or rejects quasi-periodic disturbances. For example, ILC can suppress particle-beam RF-loading transients in RF cavities for acceleration. This paper, for the first time, explains the structural causes of ``bad learning transients'' for causal and noncausal learning in terms of their eigen-system properties. This paper underscores the fundamental importance of the linear weighted-sums of the column elements of the iteration matrix in determining convergence, and the relation to the convergence of sum of squares. This paper explains how to apply the z-transform convergence criteria to causal and \underline{noncausal} learning. These criteria have an enormous advantage over the matrix formulation because the algorithm scales as $N^2$ (or smaller) versus $N^3$, where N is the length of the column vector containing the time series. Finally, the paper reminds readers that there are also wave-like (soliton) solutions of the ILC equations that may occur even when all convergence criteria are satisfied.
\end{abstract}

\section{INTRODUCTION}
Iterative Learning Control (ILC) is a method to train robots to perform repetitive tasks, or train a system to reject quasi-periodic disturbances.
ILC is concerned with iterations of a trial. A trial consists of a plant operator {\bf P} generating a time-series of values in response to an input vector {\bf d}.
The series is processed by a learning function {\bf L}. The vector ${\bf e}=({\bf I}-{\bf P}{\bf L}){\bf d}\equiv {\bf F}{\bf d}$
becomes the input for the next trial, and so on. So ILC is concerned with a sequence of series, and the convergence of that sequence.
The plant behaviour alone is stable, but the ILC may generate an input drive vector that is destructively large if the iteration scheme is unstable. 
If {\bf L} delays (lowers) or advances (lifts) the data record, learning is called causal or noncausal, respectively. "Advances" serve to pre-empt the disturbance.
In the limit of infinite vectors and matrices, there is an equivalent z-operator equation $e(z)=F(z)d(z)$ if {\bf L} is causal, and a recursion if $L(z)$ is noncausal.

The ILC concept dates back to the 1980's and achieved some degree of maturity circa 2006 as outlined in the inspirational review~[1],
which recounts conditions for asymptotic convergence (AC) based on the eigenvalues of {\bf F}, and monotonic convergence (MC)
of the error-vector norm based on the eigenvalues of ${\bf S}={\bf F}^{\rm T}{\bf F}$.
And for causal learning only, the review gives z-operator conditions (that are identical) for iteration-stability and monotonic convergence of the error-norm.
Thus it may be surprising to see ``Foundations...'' in the title of this work.
However, the  asymptotic  (and similar geometric) convergence conditions are ineffective. 
For plant operating-points  in the domain between the AC and MC conditions, extremely large transients  may occur before ultimate convergence;
so large that the plant will certainly be damaged. The review~[1] acknowledges these transients, but does not explain them.
Ref.~[2] offers an explanation of the transients, but it is unconvincing. 
In subsequent decades, ever more elaborate and sophisticated (and successful) schemes have been used avoid the learning transients.
But ``work arounds'' are not fundamental; various authors [3, 4, 5, etc] lament the incompleteness of ILC convergence theory.

{\it This work presents structural explanations of causal and non-causal learning transients}, and demonstrates
why geometric convergence of their eigen-systems of {\bf F} does not imply monotonic convergence of the error-vector norm.
{\it This work presents z-domain MC conditions for noncausal learning}, and explains how these tests may be performed using
experimental data from the plant. Further, we stress the stunning computational advantage of z-domain over eigenvalues.  

\subsection{Toeplitz matrices}
Elements of Toeplitz matrices obey the rule $F_{i,j}=F_{i+1,j+1}=f_{i-j}$. Sums of these matrices are also Toeplitz.
Special cases are the triangular forms: ``lower'' $F_{i,j}=0$ when $j>i$, and ``upper''  $F_{i,j}=0$ when $i>j$.
Pure causal/ noncausal learning matrices {\bf L} are lower/upper, respectively.
The product of upper and lower Toeplitz matrices is {\it not} Toeplitz.
The response of physical, linear systems can be described by a convolution integral with the impulse response as kernel.
The exact analogue of convolution for physical plant in discrete time is a lower Toeplitz matrix {\bf P}, where the first column is the sampled impulse response.
The iteration matrix {\bf F} is (is not)  Toeplitz for causal (nocausal) learning.

\section{MATRIX EIGEN-SYSTEMS}
We abbreviate eigenvector/eigenvalue to e-vector/e-value. Let $\lambda$ and $\sigma$ be the e-values of {\bf F} and {\bf S}, respectively.
Underlying the ``mystery'' of learning transients is that authors have focused on e-values, but not paid attention to e-vectors.
A similarity transform allows the repeated iteration to be written:
\begin{equation*}
{\bf x}_n={\bf F}^n{\bf x}_0={\bf T}\mathbf{\Lambda}^n{\bf T}^{-1}{\bf x}_0\,,
\end{equation*}
where the coupling matrix {\bf T} has columns equal the e-vectors of {\bf F}, and the diagonal matrix $\mathbf{\Lambda}$ has the e-vals of {\bf F}.
Unless {\bf F} is symmetric (which it is not), the e-values and e-vectors are complex, and e-vectors are not orthogonal.
Hence the coupling leads to constructive interference of eigen-solutions.
The Sum of Squares (SS) iterates according to:
\begin{equation*}
 {\bf x}_{n+1}^{\rm T}.{\bf x}_{n+1}={\bf x}_0^{\rm T}{\bf S}^n{\bf x}_0=
 {\bf x}_0^{\rm T}({\bf R} \mathbf{\Sigma}^n  {\bf R}^{-1}){\bf x}_0\,,
\end{equation*}
where the matrix {\bf R} has columns equal the e-vectors of {\bf S}, and the diagonal matrix $\mathbf{\Sigma}$ has the e-vals of {\bf S}.
{\bf S} is symmetric, and has real, distinct e-values and e-vectors that are orthogonal.
Modulus of all e-values  $<1$ is a sufficient condition for monotonic convergence of the vector norm, {\it only if} the e-values and e-vectors 
are real and distinct.  $\lambda$ are complex. $\sigma$ are real and distinct.
This is the root cause of transients for noncausal learning: from a complex vector basis and a 
spectrum of e-vals, it is possible to synthesize functions that initially grow and then decay. This is the analogue of the Laplace inversion integral wherein an almost arbitrary (single-sided) time function is synthesized from a spectrum of decaying exponentials.
Nevertheless, the condition largest value
$|\hat{\lambda}|\le1$ has some utility: it cuts down the domain of operating points and it's computational cost is ${\it O}(<N^2)$.

Now is the time for a revelation: a triangular Toeplitz matrix does not have an eigen-system!
The putative eigenvalue equation $({\bf F}-\lambda{\bf I}){\bf e}={\bf 0}$ has an infinite set of trivial solutions ${\bf e}={\bf 0}$ satisfied by any value 
of $\lambda$. All but one of the e-vectors of a triangular matrix are trivial zero vectors; therefore, the usual results for complete eigen-systems
(that have a full set of non-zero e-vectors) do not apply. For example, the matrix power ${\bf F}^n$ resulting from $n$ iterations
cannot be found in terms of e-vectors and e-values. As important as the condition $|\lambda|\le1$, is the region $\lambda\rightarrow 0$ which gives super-convergence for causal learning.

\subsection{Causal learning}
{\bf F} is lower Toeplitz. If the z-operator $F(z)$ is known, the elements of ${\bf F}^n$ can be found from the inverse z-transform:
\begin{equation*}
F_{i,j}^n=\frac{1}{2\pi\sqrt{-1}}\oint F(z)^n\frac{z^i}{z^j}\frac{dz}{z}\,.
\end{equation*}
Alternatively, working directly with the iteration equation ${\bf x}_{n+1}={\bf F}{\bf x}_n$, the system is solved row-by row by the method of 
forward-substitution and solving a recurrence equation for each row. The number of terms required to represent the last matrix element $F^n_{N,1}$ grows exponentially with matrix dimension $N$. Explicitly for $N=4$, the first column is:
\begin{equation*}
\begin{array}{c}
 F_1^n \\
 n F_1^{n-1} F_2  \\
 \frac{1}{2} n F_1^{n-2} \left((n-1) F_2^2+2 F_1 F_3\right)  \\
 \frac{1}{6} n F_1^{n-3} \left(\left(2-3 n+n^2\right) F_2^3+6 (n-1) F_1 F_2 F_3+6 F_1^2 F_4\right) 
\end{array} \label{eq:matrixpower}
\end{equation*}
Assuming the integer power $n$ is large, the largest single term within $F^n_{i,1}$ is $n^iF_1^{n-i}F_2^i/(i!)$. 
The competition between high powers of $n$ and the eigenvalue $F_1$ may induce apparently divergent behavior. However, the factorial in the denominator, which eventually grows faster than any single power, guarantees ultimate convergence of the series $F^n_{i,1}$ provided that $|F_1|<1$.
Thus the asymptotic behaviour depends only on $F_{i,i}=F_1$, whereas the short time-term is influenced [6] strongly by the other elements $F_j$ with $j>1$. 

\section{Z-OPERATORS}
The (unilateral) z-transform is the discrete-time version of the Laplace transform, with $z\equiv \exp(s\tau)$ and $s,z$ complex, and
$\tau$ is the sampling period. It converts an infinite  time-series into a weighted sum. z-operators manipulate infinite sums, and they provide insights to the properties of very large matrices. The operators have interesting properties, some of which we write for $F(z)$ causal. (Modifications are required for the noncausal case).
\subsubsection{Linear sums property}
This property is less well known. Let $F$ and $d$ be operator and data, respectively. Let $a$ be some particular value of $z$ larger than the circle of convergence.
\begin{eqnarray*}
d_1(z)&=&F^1(z)d_0(z)\\
\sum_{i=0}^\infty d_1[i]/a^i &=& F(a) d_0(a)=F(a)\sum_{i=0}^\infty d_0[i]/a^i\\
d_n(z)&=&F^n(z)d_0(z)
\end{eqnarray*}
\begin{eqnarray*}
\sum_{i=0}^\infty d_n[i]/a^i &=& F^n(a) d_0(a)=F^n(a)\sum_{i=0}^\infty d_0[i]/a^i\\
\sum_{i=0}^\infty d_n[i](\pm1)^i\!\!\! &=& \!\!\!F^n(\pm1) d_0(\pm1)=F^n(\pm1)\sum_{i=0}^\infty d_0[i](\pm1)^i \,. \label{eq:surprising}
\end{eqnarray*}
Here $F(\dots)$ is continuous function; and $F[\dots]$ is discrete function.
Evidently, the ratio of consecutive sums is $F(a)$; and if $|F(a)|<1$ all of these sequences decay as $n$ increases.
We may wonder what is the consequence of  $|F(z)|<1$ for all $z=\exp(i\theta)$ on the unit circle, and it is answered by Parseval's theorem:
\begin{equation*}
\sum_{i=0}^\infty d_n[i]^2=\frac{1}{\pi}\!\int_0^\pi\!\!\!\!\!\{  F^n(e^{j\theta})F^n(e^{-j\theta})\} \{ d_0(e^{j\theta})d_0(e^{-j\theta})\}d\theta\,.\label{eq:sumofsquares}
\end{equation*}
So the geometric convergence of all possible weighted sums $|F(e^{i\theta})|\le 1$ implies[6] monotonic convergence of the sum of squares (MCSS).
Notably, $F(z)$ is the transform of the first column of the causal iteration matrix. Further, if $\theta_m=m\times 2\pi/N$, where integer $m=0,1,2,\dots N$, 
then $F(z)$ is the Fourier series decomposition of the first column of {\bf F}; so all quantities needed for the MC test are physically accessible given the measured impulse response.
Further, it should be note that the two bracketing conditions $s^\pm=-1\le F(\pm1) \le 1$ are trivial to compute, and serve as preconditions: 
if either of them fail, there is no feed for deeper analysis.  

\subsection{Noncausal Learning}
Noncausal operators are those  which attempt to generate  a time series that begins before the time origin $(t=0)$. 
The physical plant {\bf P} is incapable of such an operation, and neither is the real-time control system that runs within an iteration.
However between the iterations, the stored digital data record may be manipulated at will - which is solely the domain of the learning function {\bf L}.
So a noncausal operation is made by manipulating data. 
This being so, the order of actions is important: manipulate the record, then let the plant operate on the data.
The matrices operate in the order {\bf P}.{\bf L}, and do not commute.
Noncausal learning is made by including time-advances (lifts) in the learning function.
These lifts are instituted by an upper-Toeplitz matrix. Hence, the product ${\bf P}.{\bf L}$ is lower-Hessenberg, not Toeplitz.

The z-operators discussed thus far were commuting, but what is needed are operators where the multiplication order is important.
The place to begin is with the rule for lifts on the data. The general $k$-lift operation ${\bf e}=\uparrow^k{\bf d}$ has the unilateral transform:
\begin{eqnarray*}
e(z) &=& {\cal Z}\{e[i]\}= {\cal Z}\{[d[i+k]\}\\
&=& {\cal Z}\{\uparrow^k d[i]\}=
z^k\left[d(z)-\sum_{j=0}^{k-1}\frac{d[j]}{z^j}\right]\,.
\end{eqnarray*}
\subsection{M-Term learning with a lift power series}
Suppose the learning operator is ${\bf L}=\sum_{p=0}^M\alpha_p\!\Up^p$, and iterants are related by 
${\bf x}_{n+1}=\left[{\bf I}-{\bf P}\sum_{p=0}^M\alpha_p \!\Up^p\right]{\bf x}_{n}$.\newline
The corresponding z-domain iteration is:
\begin{equation}
d_{n+1}(z)=F_M(z)d_n(z)\; +P(z)
\sum_{p=0}^M\alpha_pz^p\sum_{k=0}^{p-1}\frac{d_n[k]}{z^k}\hspace{3mm}\label{eq:single-iteration}
\end{equation}
with $F_M(z)=\left[1-P(z)\sum_{p=0}^M\alpha_pz^p\right]$. 
Starting with $n=0$, let us write the effect of two iterations:
\begin{eqnarray*}
d_2(z)&=&F_M(z)^2d_0(z) +P(z)\sum_{q=0}^M\alpha_qz^q\sum_{k=0}^{q-1}z^{-k}{d_1[k]}\\
&+&F(z)_MP(z)\sum_{p=0}^M\alpha_pz^p\sum_{k=0}^{p-1}z^{-k}d_0[k]
\end{eqnarray*}
The first term, in $F^2$, is the same as for causal learning.
The second and third terms in $P$ and $FP$, respectively,  are the cumulant effect of data loss.
Fortunately, we do not need to consider further iterations. All the information required to construct a convergence test is contained in the single
iteration Eqn.~(\ref{eq:single-iteration}).
As a general principle, the iterations do not convergence unless the sequence initiated by any single data impulse alone converges.
The data impulse $\delta(t-j\tau)$ corresponds to the sum
$d(z)=d_n[j]/z^j$ and collateral $d_n[k]\rightarrow d_n[k]\delta_{k,j}$.
Performing the summation leads to
\begin{equation}
d_{n+1}(z,j)= F_M(z)\frac{d_n[j]}{z^j}+\hspace{3.1cm}\label{eq:recursion}
\end{equation}\vspace{-3mm}
\begin{equation*}
\hspace{6mm} +P(z)\sum_{p=0}^M\alpha_pz^p
\left\{ \begin{array}{c} 
z^{-j}d_n[j]\;\;{\rm if}\;\;j\ge0\;\&\; p-j\ge 1\\ 0\;\;{\rm otherwise} \end{array}\right\}
\end{equation*}
From this equation we may either (i) find the elements $F_{i,j}$ of column $j$ of matrix {\bf F} by performing the inverse z-transform; 
or (ii) investigate the recursion as a function of $j$ and $p$; we do the latter.
For example, when $M=0$ (i.e. no lift) then $d_{n+1}=F_0(z)d_n[j]/z^j$ for all $j$;
in which case every sequence converges if $|F_0(z)|\le 1$.

\noindent
For example, when $M=1$ then  $d_{n+1}=F_0(z)d_n[0]$ if $j=0$, and  $d_{n+1}=F_1(z)d_n[j]/z^j$ if $j>0$.
Hence there are two simultaneous conditions for MC: $|F_0(z)|\le 1$ and  $|F_1(z)|\le 1$ for all $z=e^{i\theta}$.
And the equipment operating point must satisfy them both!

\noindent
Similarly, for $M=2$ there are three MC conditions for all $z=e^{i\theta}$:  $|F_0(z)|\le 1$ for $j=0$ and   $|F_1(z)|\le 1$ for $j=1$ and $F_2(z)|\le 1$ for $j\ge 2$.
The contraction to causal learning for $j<M$ is typical, and has the following interpretation and implication when $M=2$.
On the first iteration, matrix columns \#1,2 behave like a causal operation; and the remainder behave according to the double lift $\uparrow^2$.
On the second iterations, columns \#1,2,3,4 behave causally; and the remainder behave like $\uparrow^2$.
The effect slowly sweeps across the matrix; until after $N/M$ iterations the entire matrix operator behaves as ${\bf F}={\bf I}-{\bf P}$.
Thus the character of the matrix changes as the iterations progress. For general case, 
ILC system must satisfy  $M$ simultaneous z-domain monotonic convergence conditions: $|F_p(z)|\le1$ for $p=0,1,2\dots M$.
\subsection{Two convergence test paradigms}
The starting point is the measured impulse response of the physical plant for a particular operating point of the equipment.
The Zero-Order-Hold effect of the sampling has to be compensated by a lift.
From this data, we may construct the matrix operator ${\bf P}$ or samples of the z-operator $P(z=\exp[i\theta_n])$ in order to perform the
MC test. Let $0<\mu<1$ be an adjustable scalar gain.

\noindent
\underline{Matrix Operators}\newline
Construct ${\bf P}$ and ${\bf F}={\bf I}-\mu{\bf P}{\bf L}$.
Construct ${\bf S}={\bf F}^{\rm T}{\bf F}$.
Find the largest eigenvalue $\sigma$ of ${\bf S}$.
If $\sigma>1$, ILC is unstable. Consider to repeat with lower learning gain $\mu$.
Making ${\bf P}$ takes Order($N$) operations, and finding the eigenvalue takes
Order($N^3$) operations.

\noindent
\underline{Z-operators}\newline
Let $i=\sqrt{-1}$. Construct $P(z=\exp[i\theta_n])=p+i q$ from the data.
$p={\rm Re}[P]$ and $q={\rm Im}[P]$.
Construct $L(z=\exp[i\theta_n])=a+ib$ from the analytic expression for the learning scheme.
$a={\rm Re}[L]$ and $b={\rm Im}[L]$.
Construct $F(z)=1-{\mu}P(z)L(z)$.
For all values of $\theta_m$ evaluate $S(\theta_m)$\vspace{-1mm}
\begin{equation*}
=|F(z)F(z^*)|=1+2\mu(b.q-a.p)+\mu^2(a^2+b^2)(p^2+q^2)\,.
\end{equation*}
If $S(\theta_m)>1$, ILC is unstable. Consider to repeat with lower learning gain $\mu$.
Making $N$ values of $P(\theta_n)$ takes Order($N^2$) operations, and performing the test takes Order($N$) operations.
If $L(z)$ is noncausal, then the entire procedure has to be repeated for each lowered learning function until the residual $L(z)$ is causal.

The z-operator offers the advantage over matrix operators of $N^2$ versus $N^3$ computational steps.
In either case, it is important to think at the outset think about an appropriate sampling period and matrix size.
There is a huge cost to the stability analysis of choosing more samples than is 
necessary.

\section{CAUSAL 4-TERM LEARNING}
We take the plant z-transform to be $P(z)=-Az/(B-z)$, corresponding to exponential damping. The physical range of $A,B$ is $[0,1]$; but the mathematical range is
$A=[0,2]$ and $B=[-1,1]$. We take the learning operator with 3 data-lowers $L=1+v(\downarrow+\downarrow^2+\downarrow^3)$
with scalar gain $0<v\le 1$, or $L=1+v(1/z+1/z^2+1/z^3)$ in z-domain.
The geometric convergence of linear sums (GCLS) criteria, $s^\pm$, leads to the condition
\begin{equation*}
0<A<\frac{2}{1+v}\;\&\;\frac{(-2+A-A v)}{2}<B<\frac{(2-A-3 A v)}{2}\,.
\end{equation*}
The value $v=1$ leads to a (zero area) MC domain equal to the line segment $B=-1$; so is excluded.
To align with the following example, we take $v=1/3$. [$v=1/2$ behaves similarly, but has slightly stronger convergence in a slightly smaller domain $(A,B)$.]

\begin{figure}[htpb]\begin{center}
\includegraphics[height=2.5in]{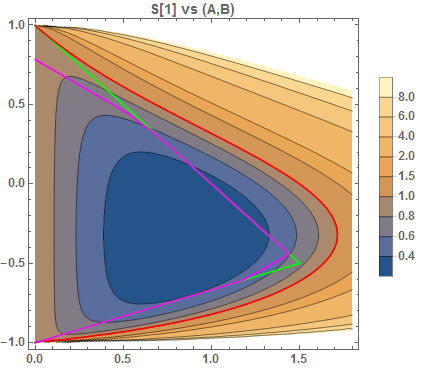}\end{center}
\caption{Contours of the ratio of squares $S[1]/1$.}
\label{fig:causal4TermMC1}
\end{figure}
\begin{figure}[htpb]
\begin{center}
\includegraphics[height=2.5in]{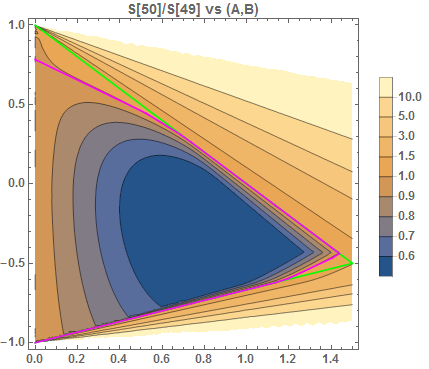}\end{center}
\caption{Contours of the ratio of squares $S[50]/S[49]$.}
\label{fig:causal4TermMC2}
\end{figure}
\begin{figure}[htpb]
\begin{center}
\includegraphics[height=2.5in]{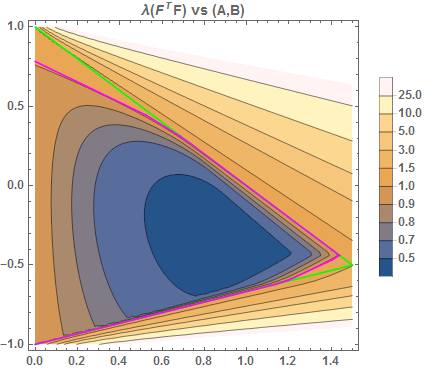}\end{center}
\caption{Largest eigenvalue $\hat{\sigma}$ of {\bf S}.}
\label{fig:causal4TermEval1}
\end{figure}
When the time-domain input data is a unit impulse, the z-domain data is the continuous spectrum $d(z)=1$.
In this case, Parseval's theorem takes a simpler form:
\begin{equation*}
S[n]=\sum_{i=0}^\infty d_n[i]^2=\frac{1}{\pi}\int_0^\pi[F(e^{i\theta})F(e^{-i\theta})]^nd\theta.
\end{equation*}
Figs.~\ref{fig:causal4TermMC1},\ref{fig:causal4TermMC2} show contours of the ratio of consecutive sum of squares, as calculated from Parseval's integral, as a function of the plant operating point $(A,B)$. 
Green lines denote the GCLS condition. Magenta lines enclose the MC domain. Red curve is the 1st iteration monotonic condition:
\begin{equation*}
 S[1]=1-2A-\frac{2}{9}A^2(4+3B)+\frac{4A^2(5+4B)}{9(1-B^2)}=1\,.
\end{equation*}
Comparison with Fig.~\ref{fig:causal4TermEval1}, the largest eigenvalue of {\bf S}, confirms the domain of convergence predicted from the MC condition.

\begin{figure}[htpb]
\begin{center}
\includegraphics[height=2.5in]{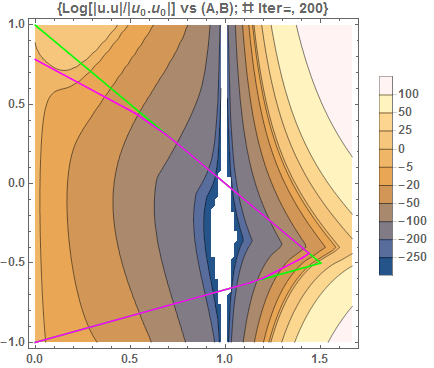}\end{center}
\caption{Contours of Log ratio of squares for 200 iterations.}
\label{fig:causal4TermRatio}
\end{figure}
\begin{figure}[htpb]
\begin{center}
\includegraphics[height=2.5in]{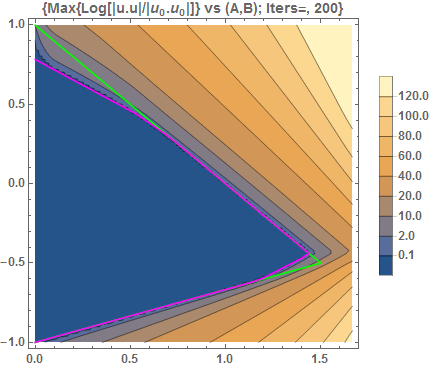}\end{center}
\caption{Log max ratio of squares for 200 iterations.}
\label{fig:causal4TermMax}
\end{figure}
Finally, we compare against direct iteration by repeated operation of the matrix {\bf F} on an input vector, for 200 iterations.
The matrix dimension is $N=100$.
The least convergent sequence is initiated by the vector ${\bf d}=(1,0,0,\dots 0)$. 
Fig.~\ref{fig:causal4TermRatio} shows the logarithm of the ratio of final to initial sum of squares, $\log_{10}\{S[200]/S[1]\}$
as function of operating point $(A,B)$.
The computer-experiment data is completely consistent with the analytic MCSS condition. 

The super-convergent effect of the eigenvalue $\lambda=1-A=0$ is clearly visible.
Fig.~\ref{fig:causal4TermMax} shows the logarithm of the ratio of largest value encountered to initial sum of squares, $\log_{10}\{\hat{S}[n]/S[1]\}$.
Comparison of Figs.~\ref{fig:causal4TermRatio},\ref{fig:causal4TermMax} betrays the learning transients: areas 
outside the MC domain where the final value is less than the maximum value.
Note, however, as predicted by the AC condition $|\lambda|=|1-A|\le 1$, equivalent to $0<A<2$,
a weak convergence has taken place across the entire domain of $(A,B)$. Although many values in Fig.~\ref{fig:causal4TermRatio} are extremely large,
nevertheless they are smaller than the values (at corresponding points) in Fig.~\ref{fig:causal4TermMax}.

The most convergent sequence is seeded by the single non-zero e-vector ${\bf d}_0=(0,0,0,\dots,1)$.
Convergence begins, and continues without interruption, across the entire domain $A=[0,2]$ and entirely independent of the value $B$.
This single e-vector iterates as ${\bf F}^n{\bf d}_0=\lambda^n{\bf d}_0$ where $\lambda=1-A$. It is this e-vector and e-value that raised
hopes~[2] of ILC convergence being independent of system dynamics; and it is the $N-1$ zero eigenvectors that dashed them.

\subsection{Negative gain}
\begin{figure}[htpb]\begin{center}
\includegraphics[height=2.5in]{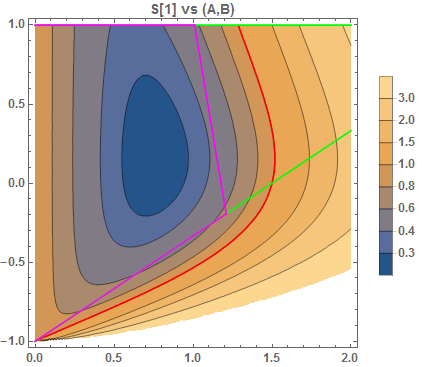}\end{center}
\caption{Contours of the ratio of squares $S[1]/1$. }
\label{fig:causal4TermNegMC1}
\end{figure}

\begin{figure}[htpb]\begin{center}
\includegraphics[height=2.5in]{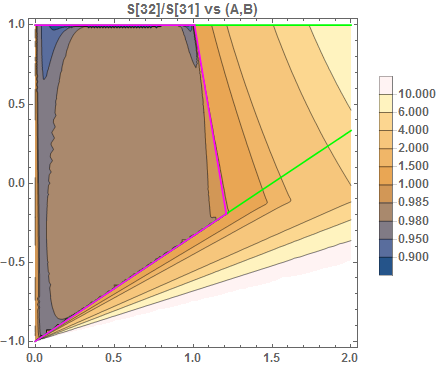}\end{center}
\caption{Contours of the ratio of squares $S[32]/S[31]$.}
\label{fig:causal4TermNegMC2}
\end{figure}

\begin{figure}[htpb]
\begin{center}
\includegraphics[height=2.5in]{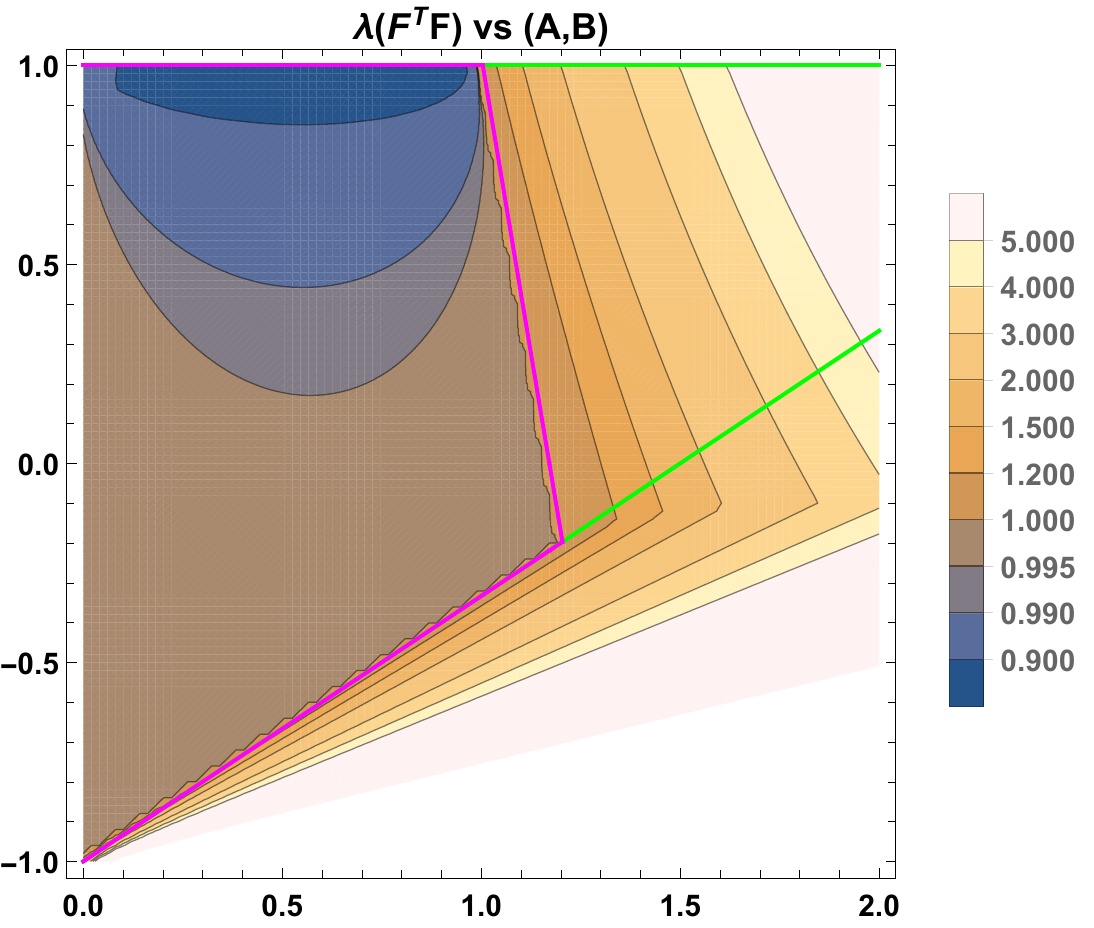}\end{center}
\caption{Largest eigenvalue $\hat{\sigma}$ of {\bf S}.}
\label{fig:causal4TermNegEval1}
\end{figure}

Suppose the z-domain learning operator is  $L=1-v(1/z+1/z^2+1/z^3)$ with $0<v\le 1$.
The GCLS criteria, $s^\pm$, leads to the conditions $0<v<1/3$ and
\begin{equation*}
0<A<2\;\&\;(-2+A+A v)<2B<(2-A+3 A v)\,.
\end{equation*}
Figs.~\ref{fig:causal4TermNegMC1},\ref{fig:causal4TermNegMC2} show instances of the progression of
the ratio of sum of squares, as may be calculated from Parseval's integral or by summing the series directly. The GCLS conditions $s^\pm$
are drawn in green, and the MCSS condition $|F(e^{i\theta}|\le1\;\forall\;\theta$ is plotted in magenta.
Red curve is the 1st iteration monotonic condition:
\begin{equation*}
S[1]=1-2A-\frac{1}{2}A^2(3+2B)+\frac{A^2(13+12B)}{4(1-B^2)}=1\,.
\end{equation*}
The convergence domain is larger than for positive gain, but convergence is weaker.
Comparison with Fig.~\ref{fig:causal4TermNegEval1}, the largest e-value of {\bf S}, confirms the domain of convergence predicted from the MC condition.
\begin{figure}[htpb]
\begin{center}
\includegraphics[height=2.5in]{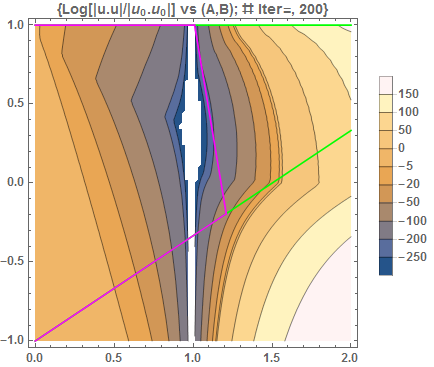}\end{center}
\caption{Contours of Log ratio of squares for 200 iterations.}
\label{fig:causal4TermNegRatio}
\end{figure}
\begin{figure}[htpb]
\begin{center}
\includegraphics[height=2.5in]{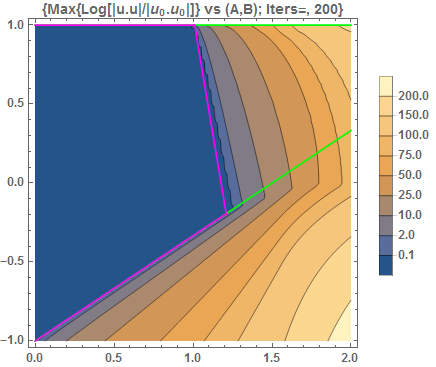}\end{center}
\caption{Log max ratio of squares for 200 iterations.}
\label{fig:causal4TermNegMax}
\end{figure}
\newpage
\section{NONCAUSAL 4-TERM LEARNING}
\begin{figure}[htpb]\begin{center}
\includegraphics[height=2.in]{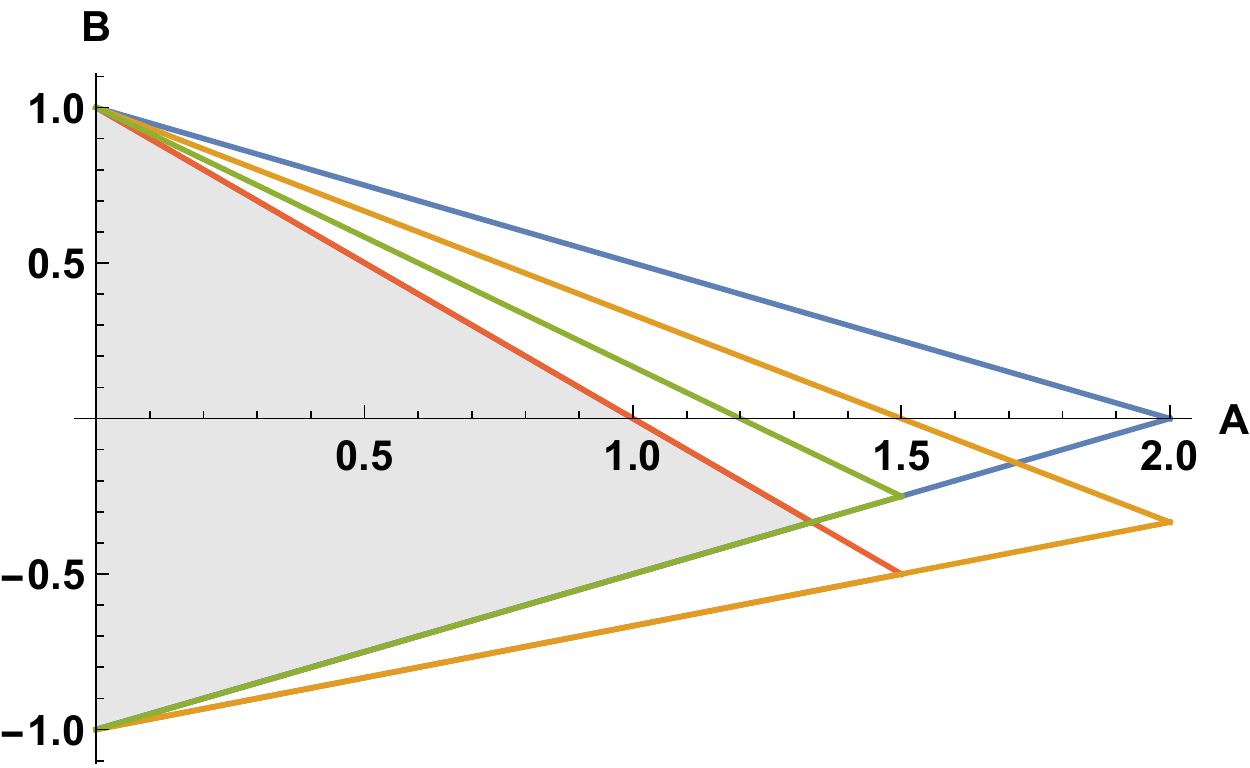}\end{center}
\caption{GCLS condition, $s_p^\pm$, for $F_0$ (blue), $F_1$ (gold), $F_2$ (olive), $F_3$ (coral). }
\label{fig:4GCLS}
\end{figure}

Consider the learning operator with three equal lifts
${\bf L}={\bf I}+v\sum_{p=1}^3\!\Up^p$, and iterants are related by 
${\bf x}_{n+1}=\left[{\bf I}-{\bf P}.{\bf L}\right]{\bf x}_{n}$.
The corresponding z-domain iteration is given by Eqn.~\ref{eq:single-iteration} with $M=3$;
with $\alpha_0=1$ and $\alpha_1=\alpha_2=\alpha_3=v$.
The sequence of sum of squares ${\bf x}_n^{\rm T}.{\bf x}_n$ will be monotonically convergent if the 
set of progressively more causal learning functions satisfy $|F_p(e^{i\theta})|\le 1$ for $p=0,1,2,3$ and $\forall\;\theta=[0,\pi]$.
\begin{eqnarray*}
F_3&=&1-P(z)[1+v(z+z^2+z^3)]\hspace{5mm}F_0=1-P(z)\\
F_2&=&1-P(z)[1+v(z+z^2)]\hspace{5mm}F_1=1-P(z)[1+vz]\,.
\end{eqnarray*}
A subset of the MCSS conditions are the bracketing GCLS conditions $s_p^\pm$; $p=0,1,2,3$. 
The four GCLS conditions are easily evaluated, and must be simultaneously satisfied.
The nett condition is: $0<v\leq 1$ and 
\begin{equation}
0<A<\frac{4}{2+3 v}\;\&\;\frac{(-2+A)}{2}<B<\frac{(2-A-3 A v)}{2}\,.\label{eq:4GCLS}
\end{equation}
To align with previous examples, set $v=1/3$.
Fig.~\ref{fig:4GCLS} shows the conditions superposed.
The inscribed area is consistent with Eqn.~\ref{eq:4GCLS}; the domain of convergence is cut down considerably.

\begin{figure}[htpb]\begin{center}
\includegraphics[height=2.in]{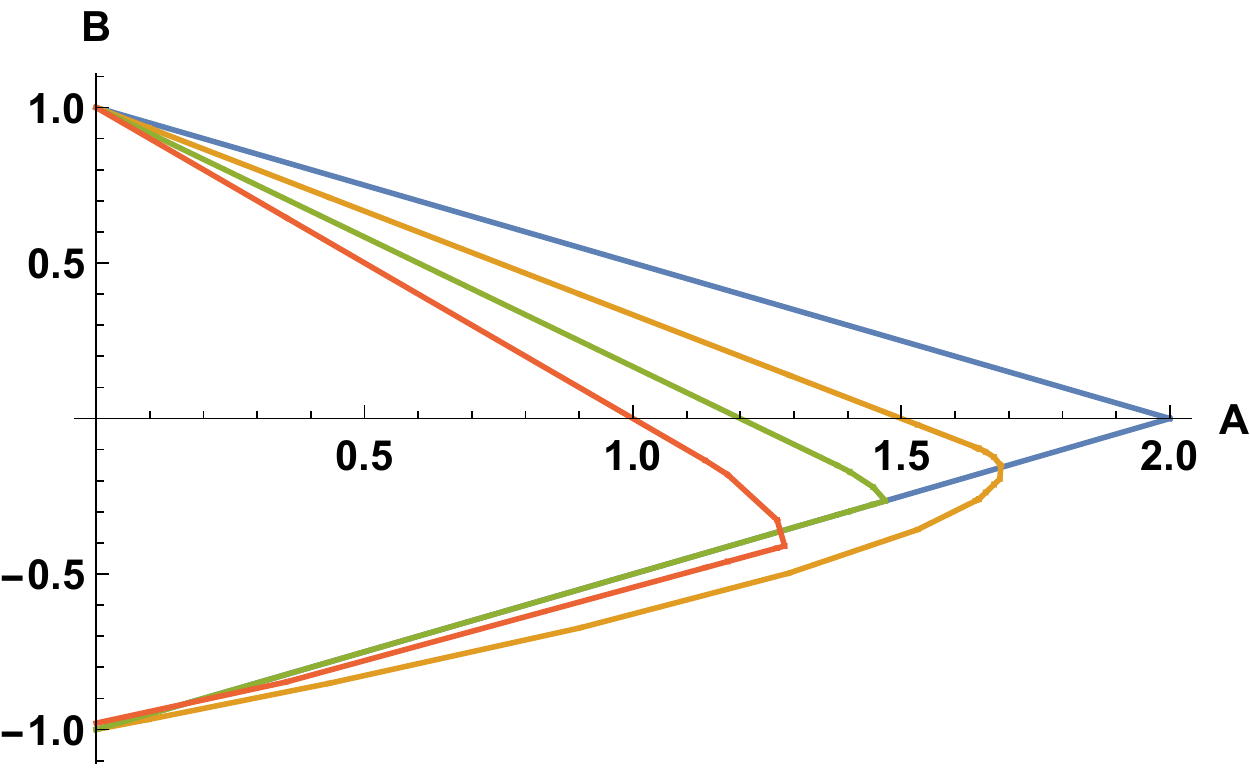}\end{center}
\caption{MCSS conditions, $|F_p|=1$, for $F_0$ (blue), $F_1$ (gold), $F_2$ (olive), $F_3$ (coral). }
\label{fig:4MCSS}
\end{figure}
The four MCSS conditions are plotted in Fig.~\ref{fig:4MCSS}. As is typical, a small number (in this case upto 12) of $\theta$ values for each $F_p$
is sufficient to define the convergence domain (the inscribed area) which is slightly smaller than that of $F_3$ alone.
The inscribed areas in Figs.~\ref{fig:4GCLS},\ref{fig:4MCSS} are very similar.

\begin{figure}[htpb]\begin{center}
\includegraphics[height=2.5in]{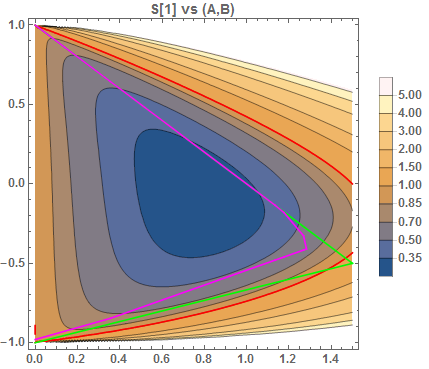}\end{center}
\caption{Contours of the ratio of squares $S[1]/1$. }
\label{fig:acausal4TermMC1}
\end{figure}

\begin{figure}[htpb]\begin{center}
\includegraphics[height=2.5in]{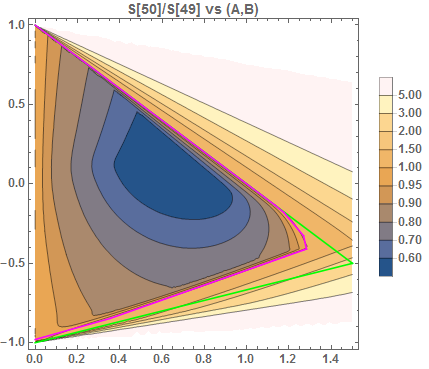}\end{center}
\caption{Contours of the ratio of squares $S[50]/S[49]$.}
\label{fig:acausal4TermMC2}\vspace{-1mm}
\end{figure}

The ratio of SS from one iteration to the next is a suitable metric for convergence.
For causal learning, there is an (almost\footnote{There is a small discrepancy due to the finite matrix size.}) exact correspondence between the vector SS and Parseval's integral.
For noncausal learning, convergence behaviour passes from $F_M$ to $F_0$, with mixed type along the way; so
the integrand has to be replaced by
{\small
\begin{eqnarray*}
\hspace{-8mm}&&\left[1-P(z) \sum _{p=0}^M z^p \alpha_p+\frac{z^j P(z) \sum _{p=0}^M z^p 
\alpha_p \sum _{k=0}^{-1+p} z^{-k} \delta _{j,k} d_n[k]}{d_n[j]}\right]\times\\
\hspace{-8mm}&&
\left[1-P(z^*)\sum _{p=0}^M (z^*)^p \alpha_p+\frac{ (z^*)^j 
P(z^*) \sum _{p=0}^M (z^*)^p \alpha_p \sum _{l=0}^{-1+p} (z^*)^{-l}
\delta _{j,l} d_n[l]}{d_n[j]}\right]
\end{eqnarray*}}
with $d_n[j]=1$ and $z=^{i\theta}$. Whichever of the $F_p$ is most restrictive will dominate
the MCSS; in this case $F_3$. 
Figs.~\ref{fig:acausal4TermMC1},\ref{fig:acausal4TermMC2} show the ratio of Parseval integrals, 
for $|F_3(z)|^{n+1}$ and $|F_3(z)|^n$  for the 1st and 50th iterations; 
and superposes the MCSS and GCLS conditions for $F_3$. 

\vspace{-2mm}

\begin{figure}\begin{center}
\includegraphics[height=2.5in]{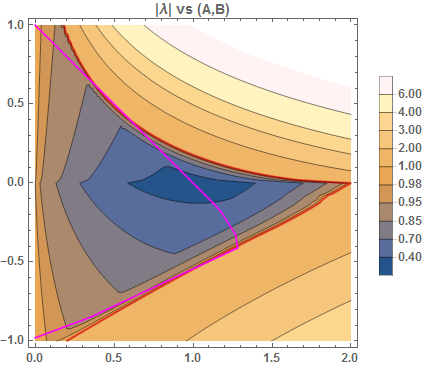}\end{center}\vspace{-1mm}
\caption{Largest eigenvalue $|\hat{\lambda}|$ of {\bf F}. Red curve is $|\lambda|=1$.}
\label{fig:acausal4TermEval1}
\vspace{-1mm}
\end{figure}

\begin{figure}\begin{center}
\includegraphics[height=2.5in]{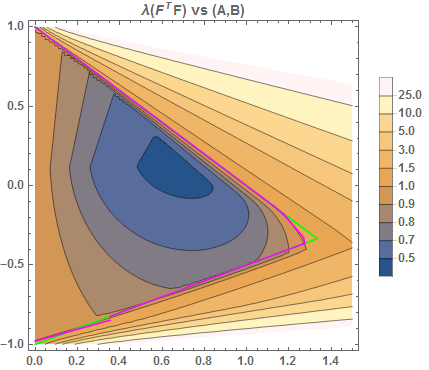}\end{center}
\caption{Largest eigenvalue $\hat{\sigma}$ of {\bf S}.}
\label{fig:acausal4TermNormEval1}
\vspace{-4mm}
\end{figure}

\begin{figure}[htpb]
\begin{center}
\includegraphics[height=2.5in]{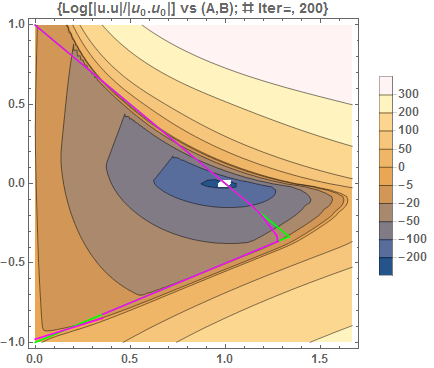}\end{center}
\caption{Log ratio of squares for 200 iterations.}
\label{fig:acausal4TermRatio}
\end{figure}

We now compare the nett MCSS domain (the inscribed area, bounded by magenta lines) with predictions based on (i) the e-vals of {\bf F}, Fig.~\ref{fig:acausal4TermEval1}; (ii) the e-vals of {\bf S}, Fig.~\ref{fig:acausal4TermNormEval1}; 
and direct iteration of {\bf F} on an input vector ${\bf d}_0$, Fig.~\ref{fig:acausal4TermRatio}.
We take the uniform seed ${\bf x}_0=(1,1,1,\dots1)$ and 200 iterations. The matrix dimension is $N=100$.

Fig.~\ref{fig:acausal4TermEval1} betrays the presence of learning transients: they will occur in the area between the red and magenta curves.
The same plot also implies that a remnant of super-convergence still exists in the neighbourhood of $A=1$. Unlike the causal case, there is a spread of e-vals; so the effect is less dramatic.

Fig.~\ref{fig:acausal4TermNormEval1} reveals the largest e-value condition $\hat{\sigma}\le1$ to 
give an identical convergence domain to the nett MCSS condition; the former consumed an hour of c.p.u. for a matrix dimension $N=50$, and the latter about a minute.

\begin{figure}[htpb]
\begin{center}
\includegraphics[height=2.5in]{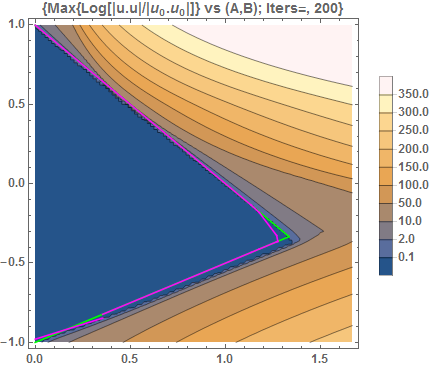}\end{center}
\caption{Log max ratio of squares for 200 iterations.}
\label{fig:acausal4TermMax}
\end{figure}

Fig.~\ref{fig:acausal4TermMax}, which shows the logarithm of the maximum value of $S[n]/S[1]$ encountered,
confirms the learning transients: immediately outside the MCSS domain, the ratio rises to $S[n]/S[1]>200$ and beyond.
It also confirms the MCSS domain has been correctly identified.

\begin{figure}[hpb]
\begin{center}
\includegraphics[height=2.5in]{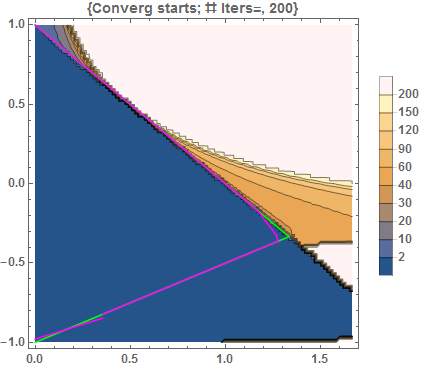}\end{center}
\caption{Contours of "when convergence starts". Pink denotes "not started during 200 iterations".}
\label{fig:acausal4TermStart}
\end{figure}

\begin{figure}[htpb]
\begin{center}
\includegraphics[height=2.5in]{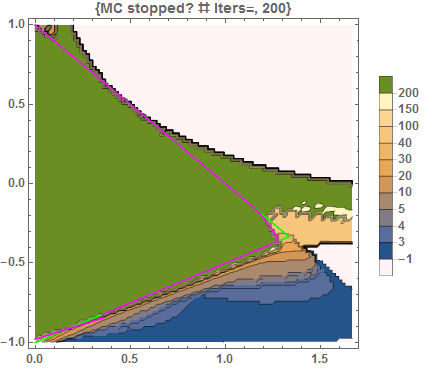}\end{center}
\caption{Contours of "when convergence stops". Pink indicates "never started". Green denotes"never stopped".}
\label{fig:acausal4TermStop}
\end{figure}

\begin{figure}[htpb]
\begin{center}
\includegraphics[height=2.5in]{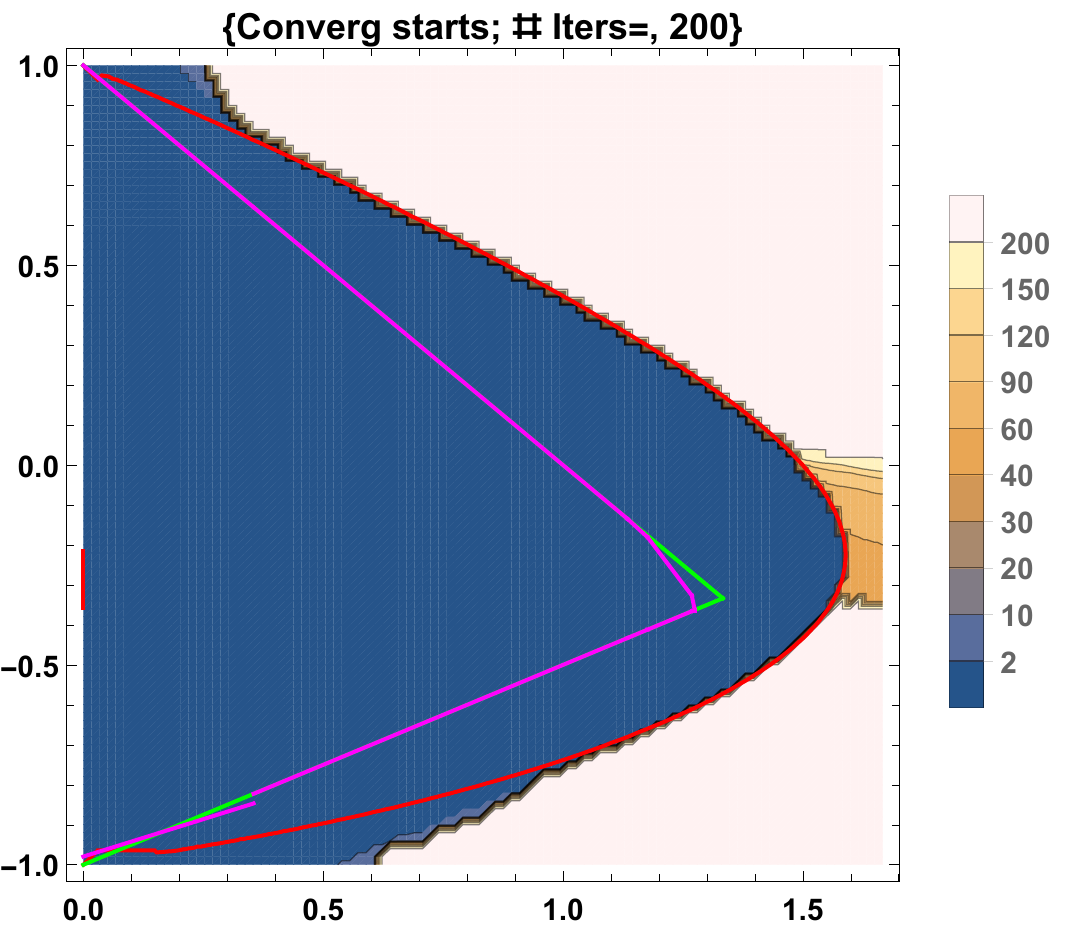}\end{center}
\caption{Contours of "when convergence starts". Pink denotes "not started during 200 iterations".}
\label{fig:acausal4TermStartN}
\end{figure}

\begin{figure}[htpb]
\begin{center}
\includegraphics[height=2.5in]{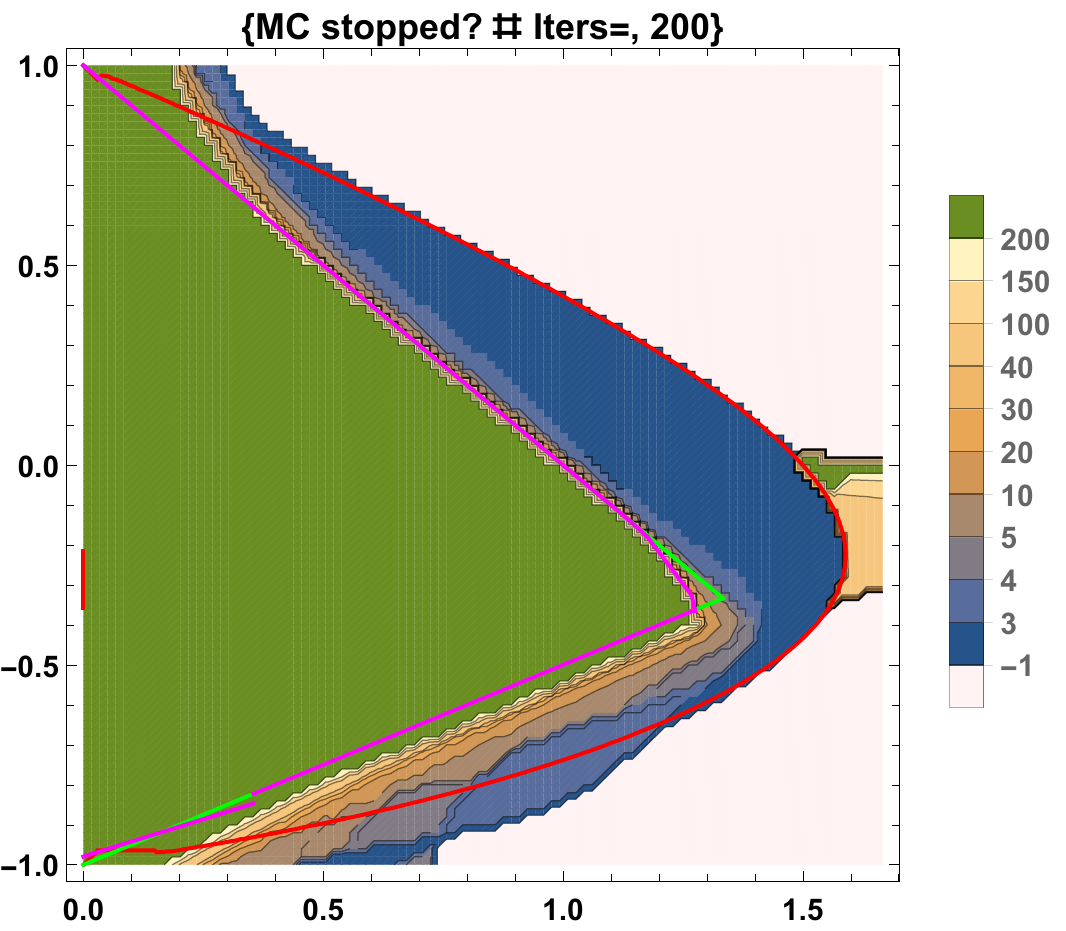}\end{center}
\caption{Contours of "when convergence stops". Pink indicates "never started". Green denotes"never stopped".}
\label{fig:acausal4TermStopN}
\end{figure}

Figs.~\ref{fig:acausal4TermStart},\ref{fig:acausal4TermStop} attempt to characteristic the progress of convergence by recording when it starts, and when it stops. There are ``false starts'' that quickly cease converging. And there are areas where convergence continues, but it is not monotonic.
Plots of this type depend on the initial seed vector ${\bf x}_0$. The two least convergent seeds are $x_0[i]=\delta_{i,N-M}$ and 
nearest neighbour $x_0[i]=\delta_{i,N-M+1}$; the progression of their iterants, Figs.~\ref{fig:acausal4TermStartN},\ref{fig:acausal4TermStopN}, are very similar; but markedly different to 
that of ${\bf x}_0=(1,1,1,\dots,1)$. The false starts are particularly evident. The red curve is the contour $S[1]=1$ calculated for $F_3(z)$.
The quantities $S[n]$ are functions of $(A,B)$ and may be calculated either by summing the series or evaluating the integral;
in this case\; $S[1]=1-(2/9)A[ (A (4 + 3 B)+$
\begin{equation*}
+ 3 (3 + B + B^2 + B^3)] + 4 A^2 (5 + 4 B)/(9 (1 - B^2)) =1\,.
\end{equation*}

Figs.~\ref{fig:acausal4TermStart}-\ref{fig:acausal4TermStopN} demonstrate a general property: the details of a convergence sequence depend
strongly on the initial seed vector. This is true both of causal and noncausal learning, and has the implication that it is risky to try and map a convergence domain
based on computer experiments - unless the seeds are specially chosen to be yield the least convergent sequences. 
Take the impulse in first position $\delta_{i,1}$ for causal, and the impulse in $N-M$ position $\delta_{i,N_M}$ for noncausal.
Better still is to perform the z-domain MCSS tests, because they are the more stringent.

\vspace{-1mm}
\vspace{-1mm}
\newpage

\section{SOLITONS}
At the outset, the eigen-system analysis of ILC presumes that the iteration index $n$ and within-trial sample-time index $k$ are the arguments of separate functions; and this implicitly excludes wave-like solutions $W(n-ck)$ where $c$ is the wave speed. Given that the ILC gain parameters are tuned for decay, ordinary waves are excluded; but not wave-packets with high-frequency carriers. To be clear, these disturbances do not appear to travel within a single trial; it is only when they are plotted in the 2-dimensional space $(n,k)$ that their motion becomes manifest. They satisfy the usual definition of a soliton wave:  a self-reinforcing wave packet that maintains its (unique) shape while it propagates at constant speed; and they persist long after all disturbances should have decayed practically to zero. The shape and group velocity must be found[7] self-consistently.
The presence of the high-frequency carrier implies they probably can be eliminated by pre-pending a low pass filter {\bf Q} to the iteration matrix:
$({\bf I}-{\bf P}{\bf L})\rightarrow {\bf Q}({\bf I}-{\bf P}{\bf L})$, but at the cost of displacing the fixed point of the mapping from zero - leading to residual error. (In graphic terms, the robot arm losses its tremor but misses the target.)

\section{CONCLUSION}
After two decades, two foundational issues in ILC are resolved. 
As promised in the introduction, this work presents structural explanations of learning transients  in causal and non-causal ILC schemes 
in terms of the properties of the eigen-systems of their respective matrix operators.
Further, this work presents z-domain MC conditions for noncausal learning, analogous to those for causal learning, but augmented
by additional terms that account for the data-loss that accrues from lift operations on data. 
The z-domain monotonic convergence criteria were compared against those calculated from matrix eigenvalues, and compared with the results of 
direct iteration of the matrix operator on initial seed vectors. All results are in perfect agreement.
Evidently, the computationally costly eigen-value calculations may be dispensed with.

%
%
\ifboolexpr{bool{jacowbiblatex}}%
	{\printbibliography}%
	{%
	
	
} 
\vfill

%
%


\end{document}